\documentclass[prb, superscriptaddress]{revtex4}
\usepackage{epsfig}
\usepackage{graphicx}
\usepackage{amsmath}
\usepackage{bm}
\usepackage{amssymb}
\usepackage{color}

\begin{document}

\title{Gap generation and phase diagram in strained graphene in a magnetic field}
\date{\today}

\author{D.O. Rybalka}
\affiliation{Department of Physics, Taras Shevchenko National Kiev University, 03022, Kiev, Ukraine}

\author{E.V. Gorbar}
\affiliation{Department of Physics, Taras Shevchenko National Kiev University, 03022, Kiev, Ukraine}
\affiliation{Bogolyubov Institute for Theoretical Physics, 03680, Kiev, Ukraine}

\author{V.P. Gusynin}
\affiliation{Bogolyubov Institute for Theoretical Physics, 03680, Kiev, Ukraine}

\begin{abstract}
The gap equation for Dirac quasiparticles in monolayer graphene in constant magnetic and
pseudomagnetic fields, where the latter is due to strain, is studied in a low-energy effective model
with contact interactions. Analyzing solutions of the gap equation, the phase diagram of the
system in the plane of pseudomagnetic and parallel magnetic fields is obtained in
the approximation of the lowest Landau level. The three quantum Hall states, ferromagnetic, antiferromagnetic, and canted antiferromagnetic, are realized in different regions of the phase
diagram. It is found that the structure of the phase diagram is sensitive to signs and
values of certain four-fermion interaction couplings which break the approximate spin-value
$SU(4)$ symmetry of the model.
\end{abstract}


\maketitle
\section{Introduction}
\label{1}

Among many remarkable properties of graphene its response to strain can be singled out as
one of efficient means to change and control the characteristics of the electronic states in
graphene. Since graphene is only one atom thick, it is easily subjected to mechanical
deformations. Various proposals to engineer strain in graphene were discussed in the literature \cite{Li,Low,Loh}. It is known that strain induces effective gauge fields \cite{Ando,Manes}
with the corresponding effective "magnetic" fields of opposite sign in valleys $K$ and $K^{\prime}$
that means that elastic deformations, unlike real magnetic field, preserve time reversal symmetry
\cite{Morozov,Balatsky}. Since time reversal symmetry is unbroken, strain induced
fields are known as pseudomagnetic fields in the literature (for a review of gauge fields in
strained graphene, see Refs.[\onlinecite{Katsnelson-review,Katsnelson-book}]).
It was proposed in Refs.[\onlinecite{Geim,Novoselov}] that a designed strain may induce uniform pseudomagnetic field, which can easily reach values exceeding 10 T.

The observation of anomalous integer quantum Hall (QH) effect  with the filling factors $\nu=\pm
4(|n|+1/2)$ ($n$ is the Landau level index) in graphene in a magnetic field \cite{QHE-experiment},
in accordance with theoretical studies in Refs.[\onlinecite{GSh,Peres}], was a milestone in graphene research as it became a direct experimental proof of the existence of gapless Dirac quasiparticles
in graphene. The four-fold degeneracy of the Landau levels in graphene is due to the $SU(4)$ symmetry connected with valley and spin. Later the plateaus $\nu=0,\pm 1, \pm 4$ in the QH effect in graphene
were observed\cite{Zhang,Jiang} in a strong magnetic field $B \geq 20\,T$. These plateaus are connected
with the magnetic field induced splitting of the $n=0$ and $n=1$ Landau levels and the degeneracy of
the lowest Landau level (LLL) is thus completely resolved.

The Landau levels related to pseudomagnetic fields\cite{Geim,Novoselov} were observed in spectroscopic
measurements.\cite{Crommie} It was pointed out\cite{Prada} that pseudomagnetic fields due to strain
can interfere in many ways with real magnetic fields. For example, the interplay of pseudomagnetic and magnetic fields in the quantum Hall regime causes backscattering in the chiral edge channels that can destroy the quantized conductance plateaus and gives rise to unconventional QH effect in strained graphene\cite{RHY} with oscillating Hall conductivity.

The gap generation in graphene in the presence of a pseudomagnetic field was studied in Ref.[\onlinecite{Herbut-coupling}]. Interestingly, it was found that unlike magnetic field which
catalyses the generation of the time reversal invariant Dirac mass, pseudomagnetic field catalyses
the generation of time-reversal symmetry breaking Haldane mass. Various competing ground states
in monolayer graphene in pseudomagnetic fields were recently studied in Ref.[\onlinecite{Sau}].
Finally, we would like to add that very strong $50-60\,T$ pseudomagnetic fields may be
realized in molecular graphene\cite{Mar}.

The interplay between different possible ground states in strained graphene in a magnetic field
represents an important unsolved problem at the moment.
In the present paper, we study a gap generation for quasiparticles in  monolayer graphene in the
presence of both constant magnetic and pseudomagnetic fields and using the model with local
four-fermion interactions considered in Refs.[\onlinecite{Kharitonov,Kharitonov-edge-states}].
Local four-fermion terms in the Hamiltonian are remnants of the interactions on the atomic
scale, and in spite being much smaller than Coulomb interaction, they play an important role in
deciding how the $SU(4)$ symmetry is broken in monolayer graphene as well as bilayer graphene.
This especially concerns the nature of the QH state with half-filled zero-energy Landau level.
We obtain the phase diagram for competing quantum Hall states in the LLL approximation
when the chemical potential is tuned to the charge neutrality point, i.e., the state with
the zero filling factor.

The paper is organized as follows. We begin by presenting in Sec.\ref{model} the model describing
low-energy quasiparticles excitations in  strained monolayer graphene in an external magnetic field
and in the presence of local four-fermion interactions. The derivation of the gap equation is given in
Sec.\ref{gap-equation-B} and its solutions are presented in Sec.\ref{general-LLL}. The phase diagram
of the system is derived and discussed in Sec.\ref{phase-diagram}. The main results are summarized in Sec.\ref{conclusions}. Appendices at the end of the paper contain technical
details and derivations used to supplement the presentation in the main text.

\section{Hamiltonian of the Model}
\label{model}

The low-energy quasiparticles excitations in graphene can be described in terms of a four-component
Dirac spinor $\Psi^{T}_{\alpha}=(\psi_{KA\alpha},\psi_{KB\alpha},\psi_{K^{\prime}B\alpha},
\psi_{K^{\prime}A\alpha})$ which combines the Bloch states with spin index $\alpha=1,2$ on the two
different sublattices $(A, B)$  and with momenta near the two nonequivalent valley points
$(K, K^{\prime})$ of the Brillouin zone. The free quasiparticle Hamiltonian has a relativistic-like
form with the Fermi velocity $v_F=10^6\,m/s$ playing the role of the speed of light
\begin{equation}
H_0=\int d^2r\,\left[\,v_F\bar{\Psi}(\gamma^1\pi_x+\gamma^2\pi_y)\Psi\,+\,\epsilon_Z\Psi^{\dagger}
\sigma_z\Psi\,\right],
\label{free-Hamiltonian}
\end{equation}
where $\bar{\Psi}=\Psi^{\dagger}\gamma^0$ is the Dirac conjugated spinor and $\mathbf{r}=(x,y)$.
The matrices $\gamma^{\nu}$ with $\nu=0,1,2$ are $4\times 4$ matrices which satisfy the
anticommutation relations of the Dirac algebra $\{\gamma^{\mu},\gamma^{\nu}\}=2g^{\mu\nu}$,
where $g^{\mu\nu}=\mbox{diag}(1,-1,-1)$ and $\mu,\nu=0,1,2$. These matrices belong
to a reducible representation of the Dirac algebra $\gamma^{\nu}=\tilde{\tau}_z\otimes(\tau_z,i\tau_y,-i\tau_x)$,
where the Pauli matrices $\tilde{\tau}_i$ and $\tau_i$ with $i=x,y,z$ act in the subspaces
of the valley $(K,K^{\prime})$ and sublattices $(A,B)$ indices, respectively.

The canonical momentum $\bm{\pi}=-i\hbar\mathbf{\nabla}+e\mathbf{A}/c+\gamma^3 \gamma^5 e
\mathbf{A}_5/c$ includes the vector potential in the Landau gauge $\mathbf{A}=(0, B_\perp x)$
corresponding to the component $\mathbf{B}_{\perp}$ of an external magnetic field $\mathbf{B}$
orthogonal to the plane of graphene, and $\mathbf{A}_5$ is the vector potential describing
the strain induced gauge fields \cite{Herbut-coupling,Jackiw,RHY,Roy}.
In the representation of the Dirac matrices that we use, $\gamma^3$ and $\gamma^5=i\gamma^0\gamma^1\gamma^2\gamma^3$ matrices equal $\gamma^3=i\tilde{\tau}_y\otimes I$
and $\gamma^5=\tilde{\tau}_x\otimes I$, where $I$ is the $2\times 2$ unit matrix. It is easy
to check that the product of these matrices $\gamma^3\gamma^5=\tilde{\tau}_z\otimes I$ is a matrix
diagonal in the subspace of valleys that ensures that the term in the canonical momentum with
the vector potential $\mathbf{A}_5$ takes opposite signs in $K$ and $K^{\prime}$ valleys.
In what follows, we will consider only the case of constant magnetic $\mathbf{B}$ and
pseudomagnetic $\mathbf{B}_5$ fields.  The pseudomagnetic field $\mathbf{B}_5$ points always
in the direction perpendicular to the plane of graphene and, therefore, is described
 by the vector potential $\mathbf{A}_5=(0,B_5x)$, where $B_5=|\mathbf{B}_5|$.

The last term in the free Hamiltonian (\ref{free-Hamiltonian}) is the Zeeman interaction
$\epsilon_Z=\mu_BB$ with $\mu_B=e\hbar/(2mc)$ being the Bohr magneton and $\sigma_z$ is Pauli
spin matrix whose eigenstates describe spin states directed along or against the magnetic field $\mathbf{B}$. Here $B=\sqrt{\mathbf{B}^2_{\perp}+\mathbf{B}^2_{||}}$ is the strength of the
magnetic field and $\mathbf{B}_{||}$ is its component parallel to the plane of graphene.
We note that the standard Zeeman interaction $\mu_B\mathbf{B}\bm{\sigma}$ can be
reduced to this form using a rotation in spin space.

The Coulomb interaction between electrons is described by the following Hamiltonian:
\begin{equation*}
H_{C} = \frac{1}{2} \int d^2r d^2r' \bar{\Psi}(\mathbf{r})\gamma^0 \Psi(\mathbf{r}) U_C(\mathbf{r}
 - \mathbf{r'}) \bar{\Psi}(\mathbf{r'})\gamma^0\Psi(\mathbf{r'}),
\end{equation*}
where $U_C(\mathbf{r})$ is the Coulomb potential. In order to simplify the
analysis, we follow the approach of Ref.[\onlinecite{graphene-QHE}] and replace the Coulomb
interaction $U_C(\mathbf{r})$ by the contact interaction $G_{int}\delta^2(\mathbf{r})$.
The Hamiltonian $H_0+H_C$ in the absence of the Zeeman term possesses a global
$SU(4)$ symmetry connected with valley and spin degrees of freedom.

Although the Coulomb interaction is the strongest interaction between electrons in graphene,
local four-fermion interactions\cite{Alicea,Aleiner} play a crucial role too. Although these
interactions are much smaller than the Coulomb one, they break, in general, the $SU(4)$ symmetry
and crucially affect the selection of the ground state of the system. A set of local valley and
sublattice asymmetric four-fermion interactions was introduced in Ref.[\onlinecite{Kharitonov}].
The $\nu=0$ quantum Hall state was studied and it was shown that the phase diagram, obtained in
the presence of generic valley and sublattice  anisotropy and the Zeeman interaction, consists
of four phases: ferromagnetic, canted antiferromagnetic (CAF), charge density wave, and Kekule
distortion. The Hamiltonian of generic local four-fermion interactions reads
\begin{equation}
H_{contact}=\frac{1}{2} \int d^2r \sum_{j,k} g_{jk}
[\bar{\Psi}(\boldsymbol{r})\gamma^0 \mathcal{T}_{jk} \Psi(\boldsymbol{r})]^2,\quad\quad
\mathcal{T}_{jk} = \tilde{\tau}_j \otimes \tau_k,
\label{four-fermion-interactions}
\end{equation}
where $j,k=x,y,z$. We do not include in $H_{contact}$ the term with $g_{00}$ as it corresponds
to the local Coulomb interaction, which has already been taken into account by $G_{int}$. In
addition, we dot not include in our model the terms with $g_{0k}$ and $g_{j0}$, which vanish in
the first order in the Coulomb interactions and arise only in the second order due to virtual
transitions to other bands\cite{Kharitonov}. The coupling constants $g_{jk}$ are not all independent.
As shown in Ref.[\onlinecite{Kharitonov}], symmetry and other considerations lead to the following equalities for nonzero constants:
\begin{equation}
g_{\bot\bot}=g_{xx}=g_{xy}=g_{yx}=g_{yy},\quad\quad g_{\bot z}=g_{xz}=g_{yz},\quad\quad
g_{z \bot}=g_{zx}=g_{zy}.
\end{equation}
Thus, totally we have four interaction coupling constants, $G_{int},g_{\bot\bot},
 g_{\bot z},g_{z \bot}$, in the considered model.
Finally, let us present ${\cal T}_{jk}$ in terms of the $\gamma$-matrices
\begin{eqnarray}
{\cal T}_{xx}&=&-i\gamma^3\gamma^2,\quad\quad {\cal T}_{xy}=i\gamma^3\gamma^1,\quad\quad
{\cal T}_{xz}=-\gamma^3\gamma^0,\nonumber\\
{\cal T}_{yx}&=&-\gamma^5\gamma^2,\quad\quad {\cal T}_{yy}=\gamma^5\gamma^1,\quad\quad
{\cal T}_{yz}=i\gamma^5\gamma^0,\nonumber\\
{\cal T}_{zx}&=&i\gamma^2,\quad\quad {\cal T}_{zy}=-i\gamma^1,\quad\quad {\cal T}_{zz}=\gamma^0.
\label{T-matrices}
\end{eqnarray}
All these matrices are normalized as ${\cal T}^2_{ij}=1.$ This presentation is useful for
the derivation of the gap equation in the next section.

\section{Gap equation}
\label{gap-equation-B}

We will solve the gap equation in the Hartree-Fock (mean-field) approximation \cite{Khveshchenko,GGMSh,Nomura,Goerbig} which is conventional and appropriate in this case.
In the subsection \ref{gap-equation-magnetic-field}, we will derive the gap equation in the
case where only real magnetic field is present. In the next subsection, we will generalize
the gap equation to the case where both magnetic and pseudomagnetic fields are present.

\subsection{Magnetic field}
\label{gap-equation-magnetic-field}

At zero temperature and in the clean limit (no impurities), the Schwinger-Dyson equation
for the quasiparticle propagator $G(u,u^{\prime})=\hbar^{-1}\langle0|T\Psi(u)\bar{\Psi}(u')
|0\rangle$ in graphene in the mean-field approximation takes the form
\begin{eqnarray}
iG^{-1}(u,u') &=& iS^{-1}(u,u') -\hbar G_{int} \gamma^0 G(u,u) \gamma^0 \delta(u-u')
 +\hbar G_{int} \gamma^0 {\rm tr}\left[\gamma^0 G(u,u)\right] \delta(u-u')\nonumber\\
 &-& \hbar \sum_{j,k} g_{jk} \{\gamma^0\mathcal{T}_{jk} G(u,u) \gamma^0\mathcal{T}_{jk}
 - \gamma^0\mathcal{T}_{jk} {\rm tr}\left[\gamma^0\mathcal{T}_{jk} G(u,u)\right] \} \delta(u-u'),
\label{gap-equation}
\end{eqnarray}
where $u=(t,\mathbf{r})$. In this subsection, we will derive the gap equation
in graphene in a magnetic field. The generalization to the case of both magnetic and pseudomagnetic
fields is rather straightforward and will be considered in the next subsection.

The inverse free propagator in the case under consideration is given by
\begin{equation}
iS^{-1}(u,u^{\prime})=[(i\hbar\partial_t-\epsilon_Z\sigma_z)\gamma^0-v_F(\bm{\pi}\cdot\bm{\gamma})]
\delta(u-u^{\prime}).
\label{inverse-free-propagator}
\end{equation}
For the full quasiparticle propagator, we will use an ansatz which is a generalization of
the ansatz used in the previous work by two of us\cite{graphene-QHE}
\begin{eqnarray}
i G^{-1}(u,u') = [ i \hbar \partial_t \gamma^0 + \mu \gamma^0 + \tilde{\mu} \gamma^0 \gamma^3
\gamma^5- v_F (\boldsymbol{\pi} \cdot \boldsymbol{\gamma})- \tilde{\Delta} + \Delta \gamma^3
\gamma^5 ]\, \delta(u-u'),
\label{ansatz}
\end{eqnarray}
where matrices $\mu,\tilde{\mu},\Delta,\tilde{\Delta}$ are defined as $\mu=\mu_\nu\sigma_\nu, \tilde{\mu}=\tilde{\mu}_\nu\sigma_\nu,\Delta=\Delta_\nu\sigma_\nu,\tilde{\Delta}=
\tilde{\Delta}_\nu\sigma_\nu$, and index $\nu$ runs the values $\nu=0,x,z$ with $\sigma_x$ and
$\sigma_z$ being Pauli spin matrices and $\sigma_0$ the unit $2\times2$ matrix [the absence of
quantities with $\sigma_y$ matrix is consistent with subsequent analysis of a gap equation].
In what follows, we consider twelve dynamically generated parameters $\mu_\nu$, $\tilde{\mu}_\nu$,
$\Delta_\nu$, and  $\tilde{\Delta}_\nu$ as constant that is consistent with our
mean-field analysis of the present model with contact interactions.

The parameters $\mu_j$ and $\tilde{\mu}_j$ with $j=x,z$ are generalized chemical potentials connected
with the QH ferromagnetism \cite{Nomura,Goerbig,Alicea,Sheng}. On the other hand, $\Delta_j$ and $\tilde{\Delta}_j$ are related to the magnetic catalysis scenario \cite{Shovkovy,Herbut,Fuchs,Ezawa}
and are Haldane and Dirac masses, respectively, and correspond to excitonic condensates (for a brief
review of the QH ferromagnetism and magnetic catalysis scenario, see Refs.[\onlinecite{Yang,GGM,Goerbig-RMP}]).
Actually, it was shown in Ref.[\onlinecite{graphene-QHE}] that the QH ferromagnetism
and magnetic catalysis scenario order parameters necessarily coexist. The physics underlying their coexistence is specific for the systems with relativistic-like quasiparticle spectrum that makes
the quantum Hall dynamics of the $SU(4)$ breakdown in graphene to be quite different from that in conventional systems with non-relativistic quasiparticle spectrum.

According to Eq.(\ref{ansatz}), the full propagator $G(u,u')$ can be written in the form
\begin{equation}
G(u,u^{\prime}) = i\langle u|\left[(i\hbar\partial_t+\mu)\gamma^0- v_F(\bm{\pi}\cdot\bm{\gamma})
+i\tilde{\mu}\gamma^1\gamma^2+i\Delta\gamma^0\gamma^1\gamma^2-\tilde{\Delta}\right]^{-1}|
u^\prime\rangle,
\label{propagator_full}
\end{equation}
where  the states $|u\rangle$ are eigenstates of the time-position operator $\hat{u}$:
$\hat{u}|u\rangle=u|u\rangle$, $\langle u|u'\rangle=\delta(u-u')$.
In Appendix \ref{A}, we derive an explicit expression for the propagator $G(u,u')$ in
the form of a sum over Landau levels.

The symmetry-breaking generalized chemical potentials and gaps $\mu_\nu$, $\tilde{\mu}_\nu$,
$\Delta_\nu,\tilde\Delta_\nu$ are related to the corresponding order parameters through the
following relationship:
\begin{equation}
\langle\bar\Psi{\cal O}_\nu\Psi\rangle=-\hbar{\rm tr}[{\cal O}_\nu G(u,u)],
\label{order_parameters}
\end{equation}
where $8\times8$ matrices ${\cal O}_\nu= \gamma^0\sigma_\nu, \gamma^0\gamma^3\gamma^5\sigma_\nu, \gamma^3\gamma^5\sigma_\nu,
\sigma_\nu$, respectively. Compared to our previous analysis,\cite{graphene-QHE,Physica-Scripta}
we included the spin matrix $\sigma_x$ in order to be able to describe the canted antiferromagnetic
state.

Since the right-hand side of the gap equation (\ref{gap-equation}) contains the full propagator at
the coincidence limit $u^{\prime}=u$, we should calculate $G(u,u^\prime)|_{u=u^\prime} = \bar{G}(u,u)$, where $\bar{G}$ is the translation invariant part of the full propagator
defined in the mixed frequency-momentum representation in Eq.(\ref{full-propagator-two-parts}).
By making use of Eqs.~(\ref{Dn-new}) and (\ref{Dn}), we find that
\begin{eqnarray}
{G}(u,u)= \int\limits_{-\infty}^{\infty}\frac{d\omega d^2k}{(2\pi)^3\hbar}\bar{G}(\omega,\mathbf{k})
=\frac{i}{2\pi l^{2}}\sum\limits_{n=0}^{\infty}\int\limits_{-\infty}^{\infty}\frac{d\omega}{2\pi\hbar}
W\frac{[P_{-}+P_{+}\theta(n-1)]}{{\cal M}-n\epsilon_B^2},
\label{interaction-new}
\end{eqnarray}
where $l=\sqrt{\hbar c/|eB_{\perp}|}$ is the magnetic length, $\epsilon_B=\sqrt{2\hbar v_F^2|eB_{\perp}|/c}\simeq424\sqrt{|B_{\perp}|[{\rm T}]}{\rm K}$ is the Landau energy scale,
$P_{\pm}$ are projectors given by Eq.(\ref{projectors}), and $W$, $M$ are matrices expressed
through $\mu$, $\tilde{\mu}$, $\Delta$, $\tilde{\Delta}$ and defined in
Eqs.(\ref{matrices-MW}) and (\ref{matrices-WM}). We note that the filling factor $\nu=2\pi l^2
\rho$ is related to the carrier imbalance $\rho=n_e-n_h$, where $n_e$ and $n_h$ are the densities of ''electrons'' and ''holes'', respectively, and $\rho$ is determined through the Green's function as
\begin{equation}
\rho = \langle0|\Psi^\dagger(u)\Psi(u)|0\rangle=-\hbar{\rm tr}[\gamma^0G(u,u)].
\label{carrier_density}
\end{equation}

Since $W$ and ${\cal M}$ contain only $\gamma^0$
and $\gamma^1\gamma^2$ Dirac matrices, it is convenient to work with eigenvectors of these matrices.
The equality $(\gamma^1\gamma^2)^2=-1$ implies that the eigenvectors $|s_{12}\rangle$ of the matrix $\gamma^1\gamma^2$ are purely imaginary
\begin{equation}
\gamma^1\gamma^2|s_{12}\rangle=is_{12}|s_{12}\rangle,\quad\quad s_{12}=\pm1\,.
\end{equation}
Similarly, since $(\gamma^0)^2=1$, the eigenvectors of the matrix $\gamma^0$ are real and given by
\begin{equation}
\gamma^0|s_0\rangle=s_0|s_0\rangle,\quad\quad s_0=\pm1.
\end{equation}
Furthermore, since $\gamma^0$ and $\gamma^1\gamma^2$ commute, we can consider states $|s_0s_{12}
\rangle$ which are simultaneously eigenvectors of $\gamma^0$ and $\gamma^1\gamma^2$  with eigenvalues
$s_0$ and $is_{12}$, respectively. The vectors $|s_0s_{12}\rangle$ form a complete basis, and
since $W$ and ${\cal M}$ contain only $\gamma^0$ and $\gamma^1\gamma^2$ matrices, the propagator
$G(u,u)$ is diagonal in the basis of $|s_{12}s_0\rangle$ vectors and is given by
\begin{eqnarray}
G(u,u;s_0,s_{12}) &=& \frac{i}{4\pi l^2}  \int_{-\infty}^{\infty} \frac{dw}{2\pi\hbar}
\sum_{n=0}^{\infty}[ s_0(\omega+m) + d][c_0 + c_j \sigma_j]^{-1}\nonumber\\
&\times& \Big( 1 + s_{12}s_\bot + [1 - s_{12} s_\bot] \theta(n-1/2)
\Big),\quad j=x,y,z.
\label{propagator-coincidence}
\end{eqnarray}
Here $s_\bot=\text{sgn}(eB_\bot)$,  the matrices $m$ and $d$ are defined in Appendix \ref{A},
and the coefficients $c_0,c_j$ are
\begin{eqnarray}
c_0&=&\omega^2 +2\omega m_0 - n \epsilon_B^2+m_\nu^2-d_\nu^2, \quad
c_y=2i s_0 (d_x m_z - d_z m_x),\\
c_x&=&2(\omega m_x + m_0 m_x - d_0 d_x),\quad c_z=2(\omega m_z + m_0 m_z - d_0 d_z),
\label{c_i}
\end{eqnarray}
where $m_\nu=\mu_\nu - s_0s_{12} \tilde{\mu}_\nu$, $d_\nu=\tilde{\Delta}_\nu + s_0s_{12} \Delta_\nu$ ($\nu=0,x,z$) and summation over dummy index $\nu$ is meant.
For strong magnetic fields, we write
\begin{equation}
G(u,u)=G_{LLL}(u,u)+G_{hLL}(u,u),
\end{equation}
where we separated the contributions of the zero Landau level, $G_{LLL}(u,u)$ with $n=0$,
and higher Landau levels, $G_{hLL}(u,u)$ with $n\ge1$, in Eq.(\ref{propagator-coincidence}).

Let us calculate first the lowest Landau level propagator $G_{LLL}(u,u)$. In order to
integrate over $\omega$ in Eq.(\ref{propagator-coincidence}), we rewrite
 the integrand by using the relation
\begin{equation}
c_0+c_j\sigma_j=[s_0(\omega+m)-d][s_0(\omega+m)+d]
\end{equation}
valid for $n=0$  and assume as usual that $\omega$ is replaced by
$\omega+i\epsilon{\rm sgn}\,\omega$ and $\epsilon\rightarrow0_{+}$.
Hence we obtain that Eq.(\ref{propagator-coincidence}) implies the following propagator
at the limit of coinciding points in the LLL approximation:
\begin{eqnarray}
G_{LLL}(u,u;s_0,s_{12})&=&\frac{i}{2\pi l^2}\frac{1+s_{12} s_\bot}{2} \int_{-\infty}^{\infty} \frac{d\omega}{2\pi\hbar}(s_0 \omega+s_0 m_\nu \sigma_\nu - d_\nu \sigma_\nu)^{-1}\nonumber\\
&=&-\frac{s_0}{4\pi\hbar l^2} \frac{1+s_{12} s_\bot}{2}
\left[\frac{(  m_i-s_0d_i)\sigma_i}{E} \theta(E-|\mu_L|)+\text{sgn}(\mu_L)\,
\theta(|\mu_L|-E)\right],\quad i=x,z
\label{propagator-coincidence-LLL}
\end{eqnarray}
where the factor $(1+s_{12} s_{\perp})/2$ with $s_{\perp}=\mbox{sgn}(eB_{\perp})$ reflects
the presence of the spin projector $P_-=(1-is_{\perp}\gamma^1\gamma^2)/2$ in the LLL contribution,
energy
\begin{equation}
E = \sqrt{( m_x-s_0d_x)^2+( m_z-s_0d_z)^2},
\label{notation-E}
\end{equation}
and we introduced the notation
\begin{equation}
\mu_L=m_0-s_0d_0=\mu_0-s_0s_{12}\tilde{\mu}_0-s_0\tilde{\Delta}_0-s_{12}\Delta_0
\label{notation-muL}
\end{equation}
for an "effective chemical potential" in the lowest Landau level.

By integrating over $\omega$, it is not difficult to check that the higher Landau
level contribution $G_{hLL}(u,u)$ diverges as $\sum^{\infty}_{n=1}\,n^{-1/2}$. Indeed,
making the change of the variable $\omega\rightarrow \sqrt{n}\epsilon_B\omega$ and taking
into account that all the dynamically generated parameters are much less than the scale
$\epsilon_B$, we find that the leading contribution at large $n\epsilon_B^2$ is given by
\begin{equation}
G_{hLL}(u,u;s_0,s_{12})\simeq\frac{1}{4\pi\hbar l^2}
\sum\limits_{n=1}^{\infty}\frac{1}{\sqrt{n}}\frac{d_\nu\sigma_\nu}{\epsilon_B}\,.
\end{equation}
This means that the right-hand side of the gap equation
(\ref{gap-equation}) diverges too. This result is the well-known artefact of using a model
with local four-fermion interactions. For a long-range interaction like, for example, the
Coulomb one considered in Ref.[\onlinecite{Physica-Scripta}], such a divergence is absent
because the gap equation contains the quasiparticle propagator at different points $u$ and
$u^{\prime}$. To proceed further, we regularize the divergence in the model under
consideration introducing a cutoff $n_{max}$ in the sum over Landau levels (a slightly
different approach was used in Ref.[\onlinecite{graphene-QHE}]), which is connected with
the ultraviolet (UV) cut-off in energy $\Lambda$ (band width) according to the relation
$n_{\max}=\Lambda^2/\epsilon^2_B$.
By using the regularization described above and retaining only the leading contribution, we find
that the higher Landau levels contribution to the propagator at the limit of coinciding points is
given by
\begin{equation}
G_{hLL}(u,u;s_0,s_{12})=\frac{\Lambda}{4\pi\hbar^3v_F^2}\,d_\nu \sigma_\nu .
\label{propagator-coincidence-higher}
\end{equation}
By combining Eqs.(\ref{propagator-coincidence-higher}) and (\ref{propagator-coincidence-LLL}),
the gap equation (\ref{gap-equation}) takes the
following final form:
\begin{eqnarray}
 s_0 m_\nu \sigma_\nu - d_\nu \sigma_\nu + s_0 s_\bot \epsilon_Z \sigma_z =& - &
 (G_{int}+g_{zz})\,G(s_0,s_{12})- 4g_{\bot \bot}\,G(s_0, -s_{12})+
 2g_{z \bot}\,G( -s_0, -s_{12})\nonumber\\
 &+& 2g_{\bot z}\,G( -s_0,s_{12})
  + g_{zz} \sum_{s_0',s_{12}'} {\rm tr}\,G(s_0',s_{12}'),
\label{eq:gapEquation}
\end{eqnarray}
where
\begin{eqnarray}
G(s_0,s_{12})&\equiv& G_{LLL}(u,u;s_0,s_{12})+G_{hLL}(u,u;s_0,s_{12})\nonumber\\
&=&\frac{-s_0}{4\pi\hbar l^2} \bigg[\frac{( m_i-s_0d_i)\sigma_i}{E}\theta(E-|\mu_L|)
+\text{sgn}\,(\mu_L)\;\theta(|\mu_L|-E)\bigg] \frac{1 + s_{12}  s_\bot}{2}
+ \frac{\Lambda}{4\pi\hbar^3v_F^2}d_\nu \sigma_\nu,
\label{G(s0,s12)}
\end{eqnarray}
and trace in the last term in Eq.(\ref{eq:gapEquation}) is taken over the Pauli spin matrices
[note that the quantities $E$ and $\mu_L$ depend on $s_0,s_{12}$ according to
Eqs.(\ref{notation-E}),(\ref{notation-muL}) and $m_\nu,d_\nu$ depend on them too]. In deriving Eq.(\ref{eq:gapEquation}), we omitted the third term on the right-hand side of Eq.(\ref{gap-equation})
which defines the Hartree contribution due to the charge density of carriers. The point is that
there are other contributions due to the charges of ions in graphene and the charges in the
substrate and gates. In view of the overall neutrality of the system, all these contributions
should cancel exactly (the Gauss law).

Finally, we would like to note that it is advantageous in deriving Eq.(\ref{eq:gapEquation})
to use $T_{jk}$ matrices given by Eq.(\ref{T-matrices}) and utilize the Dirac algebra in order
to calculate the contribution to the gap equation due to the last term in Eq.(\ref{gap-equation}).
[The matrices $T_{jk}$ have simple commutation relations with the Green's function $G(u,u)$ which
contains only $\gamma^0$ and $\gamma^1\gamma^2$ matrices (see Eq.(\ref{interaction-new})).]

\subsection{Magnetic and pseudomagnetic fields}

In this subsection, we will derive the gap equation for quasiparticles in graphene in the case
where both constant magnetic and pseudomagnetic fields are present. Then the canonical momentum
has the form
$\boldsymbol{\pi} = - i\hbar\nabla + e\boldsymbol{A}/c + \gamma^3 \gamma^5 e\boldsymbol{A}_5/c$,
where $\boldsymbol{A} = (0,B_\bot x)$ and $\boldsymbol{A}_5 = (0,B_5 x)$. Repeating the
same computations as in the previous subsection, one can show that the only difference between the
former and present cases is that the magnetic field $B_\bot$ is now replaced by the effective field $B_\bot+(i\gamma^0\gamma^1\gamma^2)B_5$ or $B_{\perp}-s_0s_{12} B_5$ in the eigenstate basis of
the matrices $\gamma^0$ and $\gamma^1\gamma^2$. Consequently, the pseudomagnetic field has opposite
signs in the $K$ and $K^{\prime}$ valleys. Therefore, in order to take into account pseudomagnetic
field, we should simply make the replacement $B_\bot \rightarrow B_\bot-s_0s_{12} B_5$ in
the corresponding equations of Subsec.\ref{gap-equation-magnetic-field} except the Zeeman energy,
where $\epsilon_Z=\mu_B\sqrt{B_\bot^2+B_\parallel^2}$, which includes the component of the magnetic
field parallel to the plane of graphene. We find it convenient in the analysis below to use the
notation $b_\parallel=B_\parallel/B_\bot$ and $b_5=B_5/B_\bot$.

Taking into account all Landau levels contributions, the gap equation for quasiparticles in graphene
in the presence of constant magnetic and pseudomagnetic fields is given by
\begin{eqnarray}
 s_0 m_\nu \sigma_\nu - d_\nu \sigma_\nu + s_0 s_\bot \epsilon_Z \sigma_z =&-&
 (G_{int}+g_{zz})\,G^{(5)}(s_0,s_{12})- 4g_{\bot \bot}\,G^{(5)}(s_0, -s_{12})\nonumber\\
 &+&2g_{z \bot}\,G^{(5)}( -s_0, -s_{12})
 +2g_{\bot z}\,G^{(5)}( -s_0,s_{12})
  + g_{zz} \sum_{s_0',s_{12}'} {\rm tr}\,G^{(5)}(s_0',s_{12}'),
\label{eq:gapEquationExact-B5}
\end{eqnarray}
where
\begin{eqnarray}
G^{(5)}(s_0,s_{12})&=&\frac{-s_0|eB_{\perp}-s_0s_{12}eB_5|}{4\pi\hbar^2 c}
\bigg[\frac{( m_i-s_0d_i)\sigma_i}{E}\theta(E-|\mu_L|)
+\text{sgn}\,(\mu_L)\;\theta(|\mu_L|-E)\bigg]\nonumber\\
 &\times&\frac{1 + s_{12}\mbox{sgn}(eB_{\perp}-s_0s_{12}eB_5)}{2}
+ \frac{\Lambda}{4\pi\hbar^3v_F^2}d_\nu \sigma_\nu.
\label{propagator-pseudomagnetic}
\end{eqnarray}
The first term on the right-hand side of Eq.(\ref{propagator-pseudomagnetic}) is the LLL contribution 
and the last one describes the contribution due to higher Landau levels.

We have found solutions of the gap  equation taking into account the contributions due to all Landau
levels. It turned out that the higher Landau levels contribution in the weak coupling regime
does not qualitatively change the results obtained in strong magnetic fields when the LLL approximation
is valid. On the other hand, the higher Landau levels contribution essentially enlarges formulas
and makes them very complicated. Therefore, in what follows, we will solve the gap equation and
present our analysis in the LLL approximation omitting the contribution due to higher Landau levels.

\section{Solutions of gap equation in the LLL approximation}
\label{general-LLL}

In this section, we consider solutions of the gap equation (\ref{eq:gapEquationExact-B5}) retaining
only the LLL contribution. Propagator (\ref{propagator-pseudomagnetic}) in the LLL approximation
contains different projectors depending on which field $B_\bot$ or $B_5$ is stronger. We will
consider both possibilities separately.

\subsection{$|B_\bot|>|B_5|$}

Let us find solutions of the gap equation (\ref{eq:gapEquationExact-B5}) in the LLL approximation 
in the case where magnetic field is stronger than pseudomagnetic field. We will use the notation of
Refs.[\onlinecite{Kharitonov,Kharitonov-edge-states}]
\begin{equation}
u_0=\frac{G_{int}}{2\pi l^2},\quad
u_z=\frac{g_{z z}}{2\pi l^2},\quad
u_\bot=\frac{g_{\bot z}}{2\pi l^2}.
\label{u-parameters}
\end{equation}
The gap equation in the LLL approximation is obtained from Eq.(\ref{eq:gapEquationExact-B5})
by replacing the full fermion Green's function $G^{(5)}$ with $G^{(5)}_{LLL}$ and multiplying 
both sides by the LLL projector $(1+s_{12}s_\perp)/2$ (for specificity, in what follows, we 
take $s_\perp=+$).
It is easy to see that in the LLL approximation twelve parameters $\mu_\nu,\tilde\mu_\nu$ and
$\Delta_\nu,\tilde\Delta_\nu$ enter in combinations $s_0(\mu_\nu-\Delta_\nu)
-(\tilde\mu_\nu-\tilde\Delta_\nu)$. Therefore, we have only six independent variables
$\mu_\nu-\Delta_\nu$ and $\tilde\mu_\nu-\tilde\Delta_\nu$. Without loss of generality we can
put $\mu_\nu=\tilde\mu_\nu=0$ so that $m_\nu=0$ and we are left only with parameters $d_\nu$.
Then the gap equation takes the following form:
\begin{eqnarray}
 - d_\nu(s_0) \sigma_\nu + s_0 \epsilon_Z \sigma_z =&-&(1-s_0 b_5)\frac{1}{2}(u_0+u_z) \bigg[
\frac{d_i(s_0)\sigma_i}{E(s_0)}\theta(E(s_0)-|d_0(s_0)|)+\;\text{sgn}(d_0(s_0))\theta(|d_0(s_0)|-E(s_0))
\bigg]\nonumber\\
&+& (1+s_0 b_5)u_\bot \bigg[\frac{d_i(-s_0)\sigma_i}{E(-s_0)}\theta(E(-s_0)-|d_0(-s_0)|)
+\;\text{sgn}(d_0(-s_0))\theta(|d_0(-s_0)|-E(-s_0))
\bigg]\nonumber\\
 & +& u_z \sum_{s'_0} (1-s'_0 b_5) \,\text{sgn}(d_0(s'_0))\, \theta(|d_0(s'_0)|-E(s'_0)),
\label{gap-equation-LLL-both}
\end{eqnarray}
where magnetic length $l=\sqrt{\hbar c/|eB_\bot|}$ and $b_5=B_5/B_{\perp}$. We find the following
solutions of the gap equation (\ref{gap-equation-LLL-both}) (solutions in the case of purely
magnetic field are considered in Appendix \ref{magnetic-LLL}):

\begin{itemize}
\item[(i)] Ferromagnetic (F) solution:

\begin{equation}
\begin{gathered}
\tilde\Delta_0=\Delta_0=\tilde\Delta_x=\Delta_x=0,\quad
\tilde{\Delta}_z = \mp \frac{b_5}{2} ( u_0+u_z-2u_\bot),\qquad
\Delta_z = \epsilon_Z \pm \frac{1}{2} (u_0+u_z+2u_\bot).
\end{gathered}
\label{F1_C}
\end{equation}
This solution exists for $|\tilde{\Delta}_z|<|\Delta_z|$. Using Eqs.(\ref{order_parameters})
and (\ref{propagator-pseudomagnetic}), it is easy to check that, for $b_5 \to 0$, this solution
is characterized by the unique order parameter $\langle\Psi^\dagger\sigma_z\Psi\rangle$ which
defines uniform magnetization. According to Eq.(\ref{ansatz}), the corresponding term $\Delta_z\sigma_z\gamma^3\gamma^5$ in the inverse full propagator describes the Haldane-type 
mass antisymmetric in spin.

\item[(ii)] Antiferromagnetic (AF) solution:

\begin{equation}
\begin{gathered}
\tilde\Delta_0=\Delta_0=\tilde\Delta_x=\Delta_x=0,\quad
\tilde{\Delta}_z = \pm \frac{1}{2} (u_0+u_z-2u_\bot),\quad
\Delta_z = \epsilon_Z \mp \frac{b_5}{2} (u_0+u_z+2u_\bot).
\end{gathered}
\label{AF1_C}
\end{equation}
This solution exists for $|\tilde{\Delta}_z|>|\Delta_z|$. If we neglect $\Delta_z$
compared to $\tilde{\Delta}_z$, the unique order parameter characterizing this solution is $\langle\bar\Psi\sigma_z\Psi\rangle$ which defines staggered magnetization. Obviously, the
term $\tilde{\Delta}_z\sigma_z$ in the inverse full propagator (\ref{ansatz}) corresponds 
to the Dirac-type mass antisymmetric in spin.

\item[(iii)] Canted antiferromagnetic (CAF) solution:
\begin{eqnarray}
\tilde\Delta_0&=&\Delta_0=\Delta_x=0,\quad
\tilde{\Delta}_x = \pm \frac{1}{2}  \sqrt{\left(1-\frac{b_5^2}{\cos^2\theta}\right)}\, (u_0+u_z-2u_\bot)\sin\theta,\nonumber\\
\Delta_z &=& \frac{\cos\theta}{2} (u_0+u_z-2u_\bot),\qquad
\tilde{\Delta}_z = - \frac{b_5}{2\cos\theta} (u_0+u_z-2u_\bot),\quad \cos\theta=-\frac{\epsilon_Z}{2u_{\perp}}.
\end{eqnarray}
This solution exists when $\tilde{\Delta}_x$ is real. For $b_5 \to 0$, this solution is
characterized by two nonzero order parameters. They are the staggered magnetization in the $x$
direction $\langle\bar\Psi\sigma_x\Psi\rangle$ and the uniform magnetization
$\langle\Psi^\dagger\sigma_z\Psi\rangle$ in the $z$ direction. It is important that the order
parameter $\langle\bar\Psi\sigma_x\Psi\rangle$ can not be transformed into $\langle\bar\Psi\sigma_z\Psi\rangle$ by means of a rotation in spin space without inducing
the uniform magnetization in the $y$ direction.

\item[(iv)] Charge density wave (CDW) solution:

\begin{equation}
\begin{gathered}
\tilde\Delta_x=\Delta_x=\tilde{\Delta}_z=0,\quad
\Delta_z=\epsilon_Z,\quad
\tilde{\Delta}_0 = \pm \frac{1}{2} (u_0-3u_z-2u_\bot),\quad
\Delta_0 = \mp \frac{b_5}{2} (u_0+u_z+2u_\bot).
\end{gathered}
\end{equation}
For $b_5 \to 0$, the order parameter which characterizes this solution is $\langle\bar\Psi\Psi\rangle$. The corresponding term $\tilde{\Delta}_0$ in the inverse full propagator (\ref{ansatz}) describes 
the Dirac mass.
\end{itemize}
It is easy to check that for $b_5 \ne 0$ pseudomagnetic field induces additional staggered 
magnetization $\langle\bar\Psi\sigma_z\Psi\rangle\sim b_5$ in the F solution,  and vice versa, 
additional uniform magnetization $\langle{\Psi^\dagger}\sigma_z\Psi\rangle\sim b_5$ in the AF 
solution. As to the CAF solution, pseudomagnetic field produces additional staggered magnetization 
in the $z$ direction. Pseudomagnetic field leads also to the generation of the Haldane mass in 
the CDW solution in addition to the Dirac mass.

The carrier density for strained graphene in the LLL approximation for $|B_\bot|>|B_5|$ is given by
\begin{equation}
\rho=\frac{1}{2\pi l^2}\sum_{s_0}(1-s_0b_5)\,{\rm sgn}(\mu_L(s_0))\,\theta(|\mu_L(s_0)|-E(s_0)),
\end{equation}
where $\mu_L(s_0)=-(s_0\tilde\Delta_0+\Delta_0)$ and $E(s_0)=\sqrt{(s_0\tilde\Delta_x+\Delta_x)^2
+(s_0\tilde\Delta_z+\Delta_z)^2}$. One can check that  the carrier density $\rho$
equals zero for the F, AF, and CAF solutions, whereas for the CDW state $\rho$ is nonzero,
thus prohibiting it. It is easy to show that for $B_5\to 0$ the density $\rho_{CDW}\sim B_5$.
Consequently, for $B_5=0$, the CDW solution becomes admissible and coincides with the corresponding
solution in the case of purely magnetic field obtained in Appendix \ref{magnetic-LLL}.
Clearly, the F, AF, CAF solutions for $B_5=0$ also reduce to those found in
Appendix \ref{magnetic-LLL}.

\subsection{$|B_\bot|<|B_5|$}

Similarly to the previous subsection it is more convenient in the case $|b_5|>1$ to use the 
following parameters rather the coupling constants $G_{int}$, $g_{zz}$, and $g_{\bot\bot}$:
\begin{equation}
u_{05}=\frac{G_{int}}{2\pi l_5^2},\quad u_{z5}=\frac{g_{z z}}{2\pi l_5^2},\quad  u_5=\frac{g_{\bot\bot}}{2\pi l_5^2},
\end{equation}
where $l_5=\sqrt{\hbar c/|eB_5|}$ and for specificity we take $eB_5>0$. The gap equation in
the LLL approximation is obtained from Eq.(\ref{eq:gapEquationExact-B5}) by replacing the
full fermion Green's function $G^{(5)}$ with $G^{(5)}_{LLL}$  and multiplying both sides by
the LLL projector $(1-s_{0})/2$. This time twelve parameters $\mu_\nu,\tilde\mu_\nu$ and
$\Delta_\nu,\tilde\Delta_\nu$ enter in combinations $(\mu_\nu+\tilde\Delta_\nu)
+s_{12}(\tilde\mu_\nu-\Delta_\nu)$. Therefore, we have only six independent variables
$\mu_\nu+\tilde\Delta_\nu$ and $\tilde\mu_\nu-\Delta_\nu$. Without loss of generality we can
put $\mu_\nu=\tilde\mu_\nu=0$ so that $m_\nu=0$, and we are left only with parameters $d_\nu$.

By making use of $m_\nu=0$ and constricting the propagator on the $s_{0}=-$ subspace, we obtain 
the following equation for parameters $d_\nu$:

\begin{eqnarray}
 - d_\nu(s_{12}) \sigma_\nu - \epsilon_Z \sigma_z =&-&(1+s_{12} b_5^{-1})\frac{1}{2}(u_{05}+u_{z5})
\bigg[\frac{d_i(s_{12})\sigma_i}{E(s_{12})}\theta(E(s_{12})-|d_0(s_{12})|)+\;\text{sgn}(d_0(s_{12}))
\theta(|d_0(s_{12})|-E(s_{12}))\bigg]\nonumber\\
&-& (1-s_{12} b_5^{-1})2u_5 \bigg[\frac{d_i(-s_{12})\sigma_i}{E(-s_{12})}\theta(E(-s_{12})-
|d_0(-s_{12})|)+\;\text{sgn}(d_0(-s_{12}))\theta(|d_0(-s_{12})|-E(-s_{12}))\bigg]\nonumber\\
& +& u_{z5} \sum_{s'_{12}} (1+s'_{12} b_5^{-1}) \,\text{sgn}(d_0(s'_{12}))\, \theta(|d_0(s'_{12})|-E(s'_{12})),
\end{eqnarray}
where  $d_\nu = \tilde{\Delta}_\nu - s_{12} \Delta_\nu$ and $E(s_{12})=\sqrt{d_x^2+d_z^2}$.

We find the following solutions of the above gap equation (solutions in the case of purely
pseudomagnetic field are considered in Appendix \ref{pseudomagnetic-LLL}):

\begin{itemize}
\item[(i)] F solution:

\begin{equation}
\begin{gathered}
\tilde\Delta_0=\Delta_0=\tilde\Delta_x=\Delta_x=0, \quad
\tilde{\Delta}_z=-\epsilon_Z\mp\frac{1}{2b_5}(u_{05}+u_{z5}+4u_5), \quad
\Delta_z=\pm\frac{1}{2}(u_{05}+u_{z5}-4u_5).
\end{gathered}
\label{F-solution-B-B-5-large}
\end{equation}
This solution exists for $|\tilde{\Delta}_z|<|\Delta_z|$.

\item[(ii)] AF solution:

\begin{equation}
\begin{gathered}
\tilde\Delta_0=\Delta_0=\tilde\Delta_x=\Delta_x=0, \quad
\tilde{\Delta}_z=-\epsilon_Z\pm\frac{1}{2}(u_{05}+u_{z5}+4u_5), \quad
\Delta_z=\mp\frac{1}{2b_5}(u_{05}+u_{z5}-4u_5).
\end{gathered}
\label{AF-B-B-5}
\end{equation}
This solution exists for $|\tilde{\Delta}_z|>|\Delta_z|$.

\item[(iii)] CAF solution:

\begin{eqnarray}
&&\tilde\Delta_0=\Delta_0=\tilde{\Delta}_x = 0,\quad
\Delta_x = \pm \sqrt{(1-z^2)\left(1-\frac{1}{b_5^{2}z^2}\right)}\,\frac{1}{2} (u_{05}+u_{z5}-4u_5),
\nonumber\\
&&\tilde{\Delta}_z = \frac{z}{2} (u_{05}+u_{z5}-4u_5),\quad
\Delta_z = - \frac{1}{2b_5z} (u_{05}+u_{z5}-4u_5),\quad z = \frac{\epsilon_Z}{4u_5}.
\label{CAF-B-B-5-large}
\end{eqnarray}
This solution exists only when $\Delta_x$ is real. According to the
analysis performed in Appendix \ref{pseudomagnetic-LLL}, the CAF solution is absent in the case
where only pseudomagnetic field is present. Therefore, the question arises what happens with
the CAF solution found here as $B_{||}=0$ and $B_{\perp} \to 0$. Obviously, $\tilde{\Delta}_z$ 
vanishes in this limit because $z$ is proportional to $\epsilon_Z$ and tends to zero. On the 
other hand, $b_5z$ does not depend on $B_{\perp}$ and, therefore, is
not sensitive to the limit $B_{\perp} \to 0$. Furthermore, the Hamiltonian of the system in
the absence of magnetic field is invariant with respect to $SU(2)$ spin rotations because
the Zeeman term vanishes in this case. Then solutions with different components $\Delta_i$
($i=x,y,z$) in $\Delta=\Delta_x\sigma_x+\Delta_y\sigma_y+\Delta_z\sigma_z$ in ansatz (\ref{ansatz}) 
but with the same $\sqrt{\Delta^2_x+\Delta^2_y+\Delta^2_z}$ are physically equivalent
because they can be easily connected by means of appropriate $SU(2)$ spin rotations. It is
not difficult to check that $\sqrt{\Delta^2_x+\Delta^2_z}$ for the CAF solution
(\ref{CAF-B-B-5-large}) equals $|\Delta_z|$ of the F solution (\ref{F-solution-B-B-5-large})
as well as $|\Delta_z|$ of the F solution (\ref{ferromagnetic-LLL-B5-1}) in the case where only pseudomagnetic field is present. Thus, the F and
CAF solutions found in this subsection are physically equivalent if magnetic field is absent.

\item[(iv)] QAH solution:

\begin{equation}
\begin{gathered}
\tilde\Delta_x=\Delta_x=\Delta_z=0,\quad
\tilde{\Delta}_z=-\epsilon_Z,\quad
\tilde{\Delta}_0 = \mp \frac{1}{2b_5} (u_{05}-3u_{z5}+4u_5),\quad
\Delta_0 = \pm \frac{1}{2}(u_{05}+u_{z5}-4u_5).
\end{gathered}
\end{equation}
\end{itemize}

Using Eq.(\ref{propagator-pseudomagnetic}), we find that the carrier density for strained graphene
in the LLL approximation equals
\begin{equation}
\rho=\frac{1}{2\pi l_5^2}\sum_{s_{12}}(1+s_{12}b_5^{-1})\,{\rm sgn}(\mu_L(s_{12}))\,\theta
\left(|\mu_L(s_{12})|-E(s_{12})\right),
\end{equation}
where $\mu_L(s_{12})=\tilde\Delta_0-s_{12}\Delta_0$ and $E(s_{12})=\sqrt{(\tilde\Delta_x-s_{12}\Delta_x)^2+(\tilde\Delta_z-s_{12}\Delta_z)^2}$.
For the F, AF, and CAF solutions, the carrier density equals zero while it is nonzero for the 
QAH solution. Thus,  the QAH state is not realized for $B_{\bot}\neq 0$. For $B_{\bot}\to 0$,
\begin{equation}
\rho_{QAH}\simeq -\frac{{\rm sgn}\Delta_0}{\pi\hbar c}\,eB_{\bot}.
\end{equation}
Therefore, for $B_{\bot}= 0$, the QAH solution becomes admissible and coincides with
the corresponding solution in Appendix \ref{pseudomagnetic-LLL}. Other solutions also reduce 
to those in Appendix \ref{pseudomagnetic-LLL} if $B_{||}=0$ and $B_{\bot}= 0$.

\section{Phase diagram of $\nu=0$ QH states in magnetic and pseudomagnetic fields}
\label{phase-diagram}

We found above several solutions of the gap equation for quasiparticles in graphene in magnetic
and pseudomagnetic fields. In order to determine which of these solutions is the ground state,
we should calculate their energy densities and then find out the phase diagram of the system. We
used the Baym-Kadanoff formalism \cite{Baym} in order to calculate the energy density $\Omega$ 
of the system:
\begin{eqnarray}
\Omega = &-&\frac{1}{4\pi\hbar c} \sum_{s_0 s_{12}} |eB_\bot-s_0 s_{12} eB_5|
\left[\left(E+ \epsilon_Z\frac{s_0 d_z- m_z}{E}\right)\theta(E-|\mu_L|)
+|d_0-s_0 m_0| \,\theta(|\mu_L|-E)\right]\nonumber\\
&\times&\frac{1 + s_{12}\mbox{sgn}(eB_{\perp}-s_0s_{12}eB_5)}{2},
\label{free-energy-density}
\end{eqnarray}
where $m_{0,z}, d_{0,z},\mu_L$, and $E$ are functions of discrete variables
$s_0,s_{12}$ (see Eqs.(\ref{notation-E}),(\ref{notation-muL}) and definitions of $m_{0,z}, d_{0,z}$
after Eq.(\ref{c_i})). The derivation of the energy density $\Omega$ is given in Appendix \ref{B}
and we retained in Eq.(\ref{free-energy-density}) only the
contribution due to the lowest Landau level. Therefore, our analysis is valid when the magnitude
of dynamically generated gaps is much less than the maximum of Landau gaps $\sqrt{2\hbar
v_F^2|eB_\perp|/c}$ or $\sqrt{2\hbar v_F^2|eB_5|/c}$.

By making use of the energy density (\ref{free-energy-density}) and the solutions found in Sec.\ref{general-LLL}, we easily calculate the following energy densities for the solutions
in the case $|b_5|<1$:
\begin{equation}
\Omega_{F}=\frac{|u_{\bot}| \mp \epsilon_Z}{\pi l^2} -C_<, \,
\Omega_{AF}=\frac{|u_{\bot}|b_5^2 \pm b_5 \epsilon_Z}{\pi l^2} - C_<,
\,\Omega_{CAF}=-\frac{\epsilon_Z^2}{4|u_{\bot}|\pi l^2} - C_<,\,
C_<=\frac{(1+b_5^2)(u_0+u_z-2u_\bot)}{4\pi l^2},
\label{energies_computed-1}
\end{equation}
where signs in the expressions for $\Omega_{F}$ and $\Omega_{AF}$ correlate with the
corresponding ones in solutions (\ref{F1_C}) and (\ref{AF1_C}). These energy densities in the
nonstrained limit equal up to a constant to the corresponding energies found in
Refs.[\onlinecite{Kharitonov,Kharitonov-edge-states}]. For $|b_5|>1$, we have
\begin{equation}
\Omega_{F}=-\frac{2u_5 b_5^{-2} \pm \epsilon_Z b_5^{-1}}{\pi l^2_5} -C_>, \,
\Omega_{AF}=-\frac{2u_5 \mp \epsilon_Z}{\pi l^2_5} -C_>, \,\Omega_{CAF}=
\frac{\epsilon_Z^2}{8u_5\pi l^2_5} -C_>,\,
C_>=\frac{(1+b_5^{-2})(u_{05}+u_{z5}-4u_5)}{4\pi l^2_5},
\label{energies_computed-2}
\end{equation}
where signs in the expressions for $\Omega_{F}$ and $\Omega_{AF}$ correlate with the
corresponding ones in solutions (\ref{F-solution-B-B-5-large}) and (\ref{AF-B-B-5}).
The terms $C_<$ and $C_>$ describe a common shift in energy density for three
solutions, therefore, they are not important for determining the ground state. Hence the
energy density depends only on two essential four-fermion coupling constants $g_{\perp z}$ and
$g_{\perp\perp}$ (related to $u_\perp$ and $u_5$, respectively).

Using the energy densities  Eqs.(\ref{energies_computed-1}) and
(\ref{energies_computed-2}), we find the phase diagram of the $\nu=0$ QH states in graphene in
constant magnetic and pseudomagnetic fields. More precisely, we fix the value of
perpendicular magnetic field $B_\perp$ and  study the phase diagram in the plane $(b_{||}, b_5)$,
where $b_{||}=B_{||}/B_\bot$ and $b_5=B_5/B_\bot$. In our analysis, following Kharitonov\cite{Kharitonov}, we assume that $u_\bot<0$. Then for $2|u_\bot|>\epsilon_Z$
(recall that $\epsilon_Z=\mu_B|B_\perp|\sqrt{1+b_{||}^2}$), we find that
the phase diagram depends on coupling constant $g_{\perp\perp}$.

\begin{figure}
\includegraphics[width=6cm]{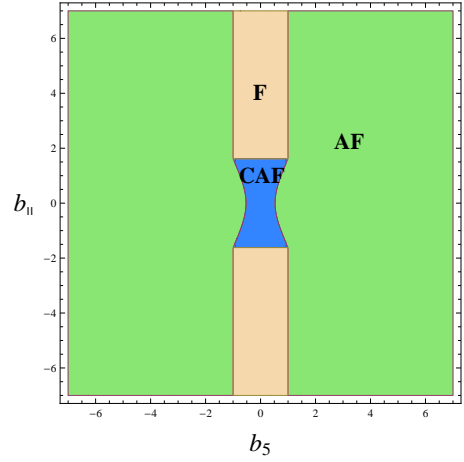}
\caption{(Color online) The phase diagram of the $\nu=0$ QH states in graphene in the plane
$(b_{||},b_5)$ for $g_{\bot\bot}>0$. The blue color represents the CAF state, green  - AF state,
and yellow  - F state. This diagram is obtained for $B_\bot=20\,T$,  $u_\bot=-13K$, $2|e|g_{\bot\bot}/(\mu_B\pi\hbar c)=2.8$, and
$\epsilon_Z=13.4 \sqrt{1+b_{||}^2}\,K$.}
\label{fig:positive-u-5}
\end{figure}

The simplest phase diagram is realized for
positive $g_{\perp\perp}$, which we plot in Fig.\ref{fig:positive-u-5}. The CAF solution is found
to be the ground state of the system in the region defined by $|b_5|<\epsilon_Z(b_{||})/2|u_\bot|<1$,
where this solution exists. The side borders of the region with the CAF solution are
determined from the condition $|b_5|=\cos\theta=\epsilon_Z(b_{||})/(2|u_\bot|)$.
This makes the central blue regions, where the CAF state is realized, look like biconcave lens.
The top and bottom borders of the region with the CAF solution are determined from the condition $\cos\theta=\epsilon_Z(b_{||})/(2|u_\bot|)=1$.

\begin{figure}[t]
\includegraphics[width=0.35\columnwidth]{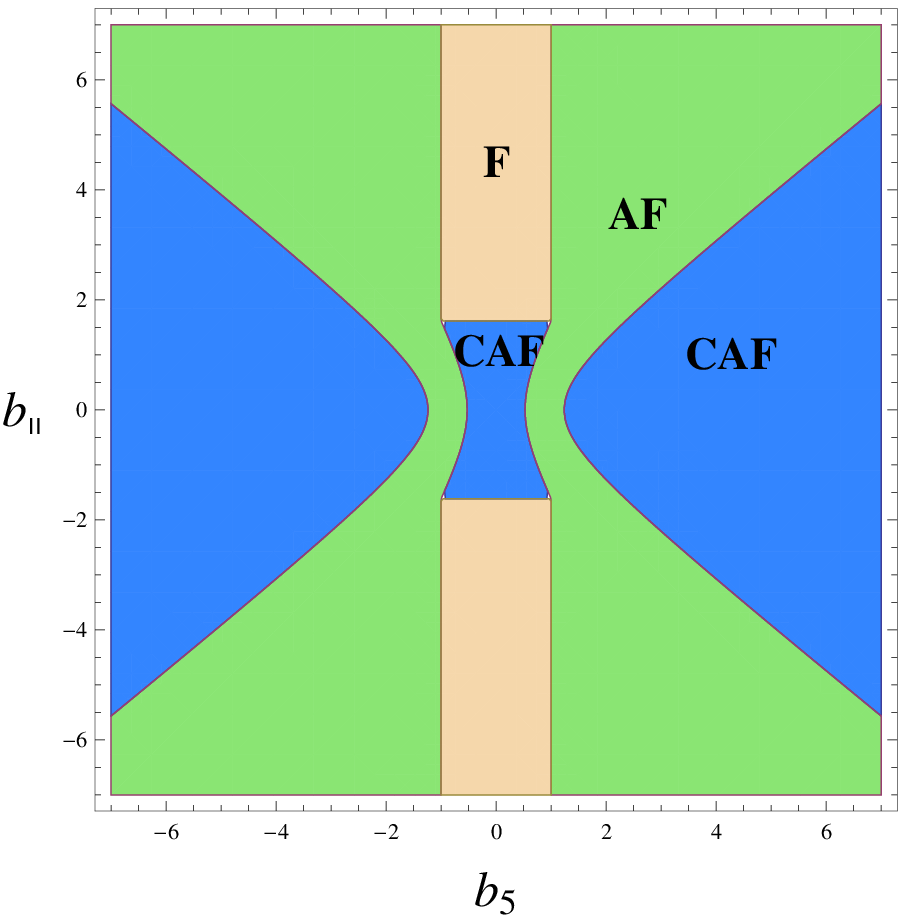}\quad
\includegraphics[width=0.35\columnwidth]{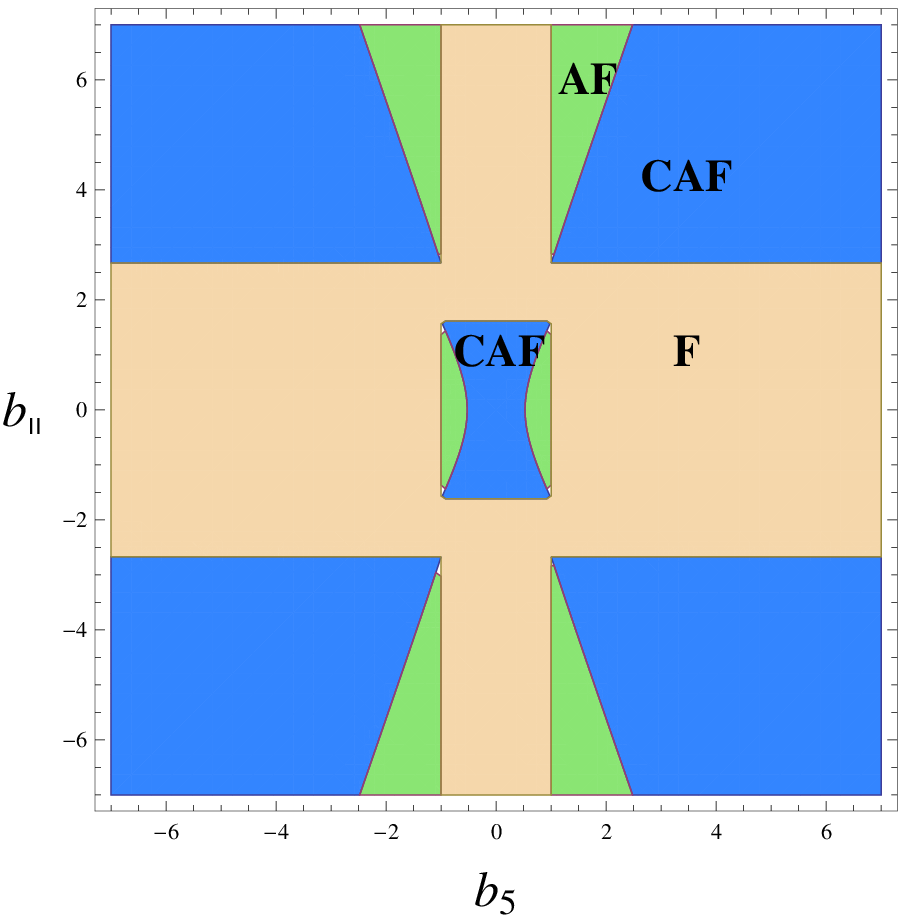}
\caption{(Color online) The phase diagram of the $\nu=0$ QH states in graphene in the plane
$(b_{||},b_5)$ for $-\mu_B\pi\hbar c/(2|e|)<g_{\bot\bot}<0$ (left panel) and $g_{\bot\bot}<
-\mu_B\pi\hbar c/(2|e|)$ (right panel). The blue color represents the CAF state, green - AF state,
and yellow - F state. These diagrams similarly to Fig.\ref{fig:positive-u-5} are obtained for
$B_\bot=20\,T$,  $u_\bot=-13\,K$, and $\epsilon_Z=13.4 \sqrt{1+b_{||}^2}\,K$. For the left panel, $2|e|g_{\bot\bot}/(\mu_B\pi\hbar c)=-0.8$,
whereas $2|e|g_{\bot\bot}/(\mu_B\pi\hbar c)=-2.8$ for the right panel. }
\label{fig:negative-u-5}
\end{figure}
The ferromagnetic solution with the minus sign in Eq.(\ref{energies_computed-1}) is
the ground state for $|b_{||}|>b_{||}^{cr}$ and $|b_5|<1$, where $b^{cr}_{||}$ is  determined by
$\epsilon_Z(b_{||}^{cr})=2|u_\bot|$, and the AF solution is the most preferable solution for the
regions  $|b_5|>1$ and $|b_{||}|<b_{||}^{cr}$,
$1>|b_5|>\epsilon_Z(b_{||})/2|u_\bot|$. As to the AF solution, plus and minus
sign for it in Eq.(\ref{energies_computed-1}) is preferable  for $b_5<0$ and $b_5>0$, respectively.
The regions of the CAF, F, and AF states are shown in Fig.\ref{fig:positive-u-5} by blue, yellow,
and green colors, respectively.

The phase diagram of the system is somewhat more complicated for negative $g_{\perp\perp}$.
There are two qualitatively different cases, namely, $-\mu_B\pi\hbar c/(2|e|)<g_{\bot\bot}<0$ and $g_{\bot\bot}<-\mu_B\pi\hbar c/(2|e|)$. The corresponding phase diagrams are plotted in the left and
right panels of Fig.\ref{fig:negative-u-5}, respectively, where like in Fig.\ref{fig:positive-u-5}
the blue color represents the CAF state, yellow - F state, and green - AF state.
Note the symmetry of the phase diagrams with respect to the change $b_5\to-b_5$ and $b_{||}
\to -b_{||}$. The main difference of the phase diagram in the left panel of Fig.\ref{fig:negative-u-5} compared to that in Fig.\ref{fig:positive-u-5} is the appearance of two additional regions, where 
the CAF state is realized as the ground state of the system. The reason
for this is that the terms with $u_5$ in the energy densities of the AF and CAF solutions (\ref{energies_computed-2}) for $|b_5|>1$ are of different sign. Therefore, for $g_{\perp\perp}>0$,
the energy density of the AF state is smaller than that of the CAF state. The situation
changes for $g_{\perp\perp}<0$, where the energy density of the CAF state is smaller than that of
the AF state for sufficiently small $b_{||}$. As $b_{||}$ increases, the CAF state ceases to exist for
$\epsilon_Z>|4u_5|$ and transforms into the AF state.  The side borders of two regions
with the CAF solution in the left panel in Fig.\ref{fig:negative-u-5} are determined by the condition
$(\mu_B\pi\hbar c/(2|e||g_{\perp\perp}|)) \sqrt{1+b_{||}^2}=|b_5|$.

For $g_{\bot\bot}<-\mu_B\pi\hbar c/(2|e|)$, the phase diagram becomes even more complicated.
Though in this case for $|b_5|>1$ the CAF solution is also preferable in every point where it 
exists, this region is now separated in four detached parts. The CAF-AF borders are analogously 
determined by the condition $(\mu_B\pi\hbar c/(2|e||g_{\perp\perp}|))\sqrt{1+b_{||}^2}=|b_5|$. 
The CAF-F borders are constant in $b_{||}$ and determined from the condition $(2|e||g_{\perp\perp}|/(\mu_B\pi\hbar c))=
\sqrt{1+b_{||}^2}$. The F solution is more preferable than the AF solution for small $b_{||}$ 
and $|b_5|>1$ as is seen from the energy densities in Eq.(\ref{energies_computed-2}).

It is instructive to compare the obtained results with those in Refs.[\onlinecite{Kharitonov,Kharitonov-edge-states}]  where purely magnetic field was considered.
The case of purely magnetic field corresponds to the line $b_5=0$ in our phase diagrams. According
to Figs.\ref{fig:positive-u-5} and \ref{fig:negative-u-5}, there is the continuous phase transition
from the CAF state to the F state as parallel magnetic field $B_{||}$ increases that  agrees with Kharitonov's findings.

\section{Summary and discussions}
\label{conclusions}

In the present work, we studied the $\nu=0$ quantum Hall states in strained monolayer graphene
under tilted external magnetic field in the presence of local Coulomb and other local four-fermion
interactions, which, in general, break the approximate spin-valley SU(4) symmetry of the
low-energy electron Hamiltonian of graphene. Solving the gap equation in the LLL
approximation, we found a rich phase diagram where different QH states, the ferromagnetic, antiferromagnetic, and canted antiferromagnetic,  compete when we change parallel
magnetic field $B_{||}$ and pseudomagnetic field $B_5$ (we keep perpendicular magnetic field
$B_\perp$ fixed and rather large). For zero strain, these states are characterized by the
nontrivial order parameters $\Psi^\dagger\sigma_z\Psi$, $\bar\Psi\sigma_z\Psi$, and
$\bar\Psi\sigma_x\Psi$ and $\Psi^\dagger\sigma_z\Psi$, respectively. In the presence of
pseudomagnetic field these order parameters gain an admixture of uniform staggered magnetization
or uniform magnetization in the $z$ direction.

Assuming that the strength of the Coulomb interaction is larger than all other local interactions,
we found an essential dependence of the phase diagram only
on two four-fermion couplings $g_{\perp z}$ and $g_{\perp\perp}$, which appear in the low-energy
effective Hamiltonian (\ref{four-fermion-interactions}). Our main results are accumulated in Figs.\ref{fig:positive-u-5} and \ref{fig:negative-u-5}. For negative coupling $g_{\perp z}$
we found that the canted antiferromagnetic state is always preferable in
the center of all figures, i.e., for not too large values of $B_{||}$ and $B_5$. When $B_{||}$
and $B_5$ increase, the ferromagnetic, antiferromagnetic, and canted antiferromagnetic states are
realized as the ground state of the system  depending on the values of $B_{||}$, $B_5$.

The particular case of purely magnetic field corresponds to the line $B_5=0$ in
our Figs.1 and 2 where there is a phase transition from the CAF state to the F state as
parallel magnetic field $B_{||}$ increases that agrees with Kharitonov`s findings.\cite{Kharitonov,Kharitonov-edge-states}.
In the case of purely pseudomagnetic field considered in Appendix \ref{pseudomagnetic-LLL}, we find
that the F and QAH solutions for $g_{\bot\bot}<0$ have the lowest and equal energy densities.
Since these solutions are described by Haldane-type masses, our results agree with the findings in
Ref.[\onlinecite{Herbut-coupling}], where it was shown that pseudomagnetic field catalyses
the generation of the Haldane mass. On the hand, for $g_{\bot\bot}>0$, the AF state with Dirac-type
mass is the most preferable. These results and those obtained in Sec.\ref{phase-diagram} show that
the structure of the phase diagrams in Figs. \ref{fig:positive-u-5} and \ref{fig:negative-u-5} is
sensitive to the sign and the strength of four-fermion coupling $g_{\perp\perp}$ as well as coupling
$g_{\perp z}$.

In our analysis, we ignored boundaries of graphene. However, it is known that they
may play an essential role in strained graphene. A novel magnetic ground state, where
the Neel and ferromagnetic orders coexist, was reported for the Hubbard Hamiltonian in strained
graphene in Ref.[\onlinecite{edge-compensated}]. Whereas the Neel order takes the same sign 
through the entire system, the magnetization at the boundary takes the opposite sign from the bulk.
Since the total magnetization vanishes, the magnetic ground state is edge-compensated
antiferromagnet.

In conclusion, strained graphene in a magnetic field provides a unique opportunity to
observe various symmetry breaking phases. Further progress in achieving  strain induced
pseudomagnetic fields, especially uniform pseudomagnetic fields, might allow one to probe 
experimentally the obtained phase diagram. As for the future, it would be interesting to extend 
the results of the present paper beyond the neutral point with the filling factor $\nu=0$ and 
describe the quantum Hall states with other filling factors.

\section{Acknowledgments}
We are grateful to S.G. Sharapov for useful suggestions and discussions.
The work of E.V.G. and V.P.G. was supported partially by the European IRSES Grant No. SIMTECH No. 246937
and by the Program of Fundamental Research of the  Physics and Astronomy Division of the NAS of Ukraine.

\appendix

\section{Quasiparticle propagator: Expansion over LLs}
\label{A}

For the Fourier transform in time of
the full propagator $G(u,u^\prime)$ in Eq.~(\ref{propagator_full}) we can write
\begin{eqnarray}
G(\omega,\mathbf{r},\mathbf{r}^{\prime}) &=& i\langle \mathbf{r}|\left[(\omega+m)\gamma^0- v_F(\bm{\pi}\cdot\bm{\gamma})-d\right]^{-1}|\mathbf{r}^\prime\rangle
=i\langle \mathbf{r}| \left[(\omega+m)\gamma^0 - v_F(\bm{\pi}\cdot\bm{\gamma})
+d\right]\nonumber\\
&\times& \left[\left((\omega+m)\gamma^0 -v_F(\bm{\pi}\cdot\bm{\gamma})-d\right)
\left((\omega+m)\gamma^0 -v_F(\bm{\pi}\cdot\bm{\gamma})
+d\right)\right]^{-1}|\mathbf{r}^\prime\rangle
\nonumber\\
&=&i\left[W-v_F(\bm{\pi}_{r}\cdot{\pmb\gamma})\right]\langle\mathbf{r} |
\left({\cal M}-v_F^{2}\bm{\pi}^{2}-ieB_{\perp}(\hbar v_F^{2}/c)\gamma^1\gamma^2\right)^{-1}
| \mathbf{r}^{\prime}\rangle ,
\label{propagator1}
\end{eqnarray}
where $m=\mu+i\gamma^0\gamma^1\gamma^2\tilde{\mu}$, $d=\tilde{\Delta}-i\gamma^0\gamma^1\gamma^2\Delta$, 
the matrices $\mu,\tilde{\mu},\Delta,\tilde{\Delta}$ are defined after Eq.(\ref{ansatz}), and
matrices $W$ and ${\cal M}$ are
\begin{eqnarray}
W&=&(\omega+m)\gamma^{0}+d,\quad m=m_\nu\sigma_\nu,\quad d=d_\nu\sigma_\nu,\quad \nu=0,x,z,
\label{matrices-MW}\\
{\cal M} &=& c+c_i\sigma_i,\quad i=x,y,z,
\label{matrices-WM}
\end{eqnarray}
with
\begin{eqnarray}
&&c=\omega^2+2\omega m_0+m_\nu^2-d_\nu^2,\quad c_x=2(\omega m_x+m_0m_x-d_0d_x),\nonumber\\
&&c_y=2i\gamma^0(d_x m_z-d_z m_x),\quad c_z=2(\omega m_z+m_0m_z-d_0d_z).
\end{eqnarray}
Our aim is to find an expression for the propagator (\ref{propagator1}) as an expansion over LLs
(we follow below the consideration in Appendix A in Ref.[\onlinecite{graphene-QHE}]).
The operator $\bm{\pi}^{2}$ has well known eigenvalues $(2n+1)\hbar|eB_{\perp}|/c$ with $n=0,1,2,\dots$
and its normalized wave functions in the Landau gauge $\mathbf{A}=(0,B_{\perp}x)$ are
\begin{eqnarray}
\psi_{np}(\mathbf{r})=\frac{1}{\sqrt{2\pi l}}\frac{1}{\sqrt{2^nn!\sqrt{\pi}}}
H_n\left(\frac{x}{l}+pl\right)e^{-\frac{1}{2l^2}(x+pl^2)^2} e^{ipy},
\end{eqnarray}
where $H_{n}(x)$ are the Hermite polynomials and $l=\sqrt{\hbar c/|eB_{\perp}|}$ is the
magnetic length. These wave functions satisfy the conditions of normalizability
\begin{eqnarray}
\int d^{2}{r}\psi^{*}_{np}(\mathbf{r})\psi_{n^{\prime}p^{\prime}}(\mathbf{r})=\delta_{nn^{\prime}}
\delta(p-p^{\prime}),
\end{eqnarray}
and completeness
\begin{eqnarray}
\sum\limits_{n=0}^{\infty}\int\limits_{-\infty}^{\infty} dp\psi^{*}_{np}(\mathbf{r})
\psi_{np}(\mathbf{r}^{\prime})
=\delta(\mathbf{r}-\mathbf{r}^{\prime}).
\label{completeness}
\end{eqnarray}
Using the spectral expansion of the unit operator (\ref{completeness}), we obtain
\begin{eqnarray}
&&\langle\mathbf{r}|\left(
{\cal M}-v_F^{2}\bm{\pi}^{2}-ie B_{\perp}(\hbar v_F^{2}/c)\gamma^1\gamma^2\right)^{-1}|
\mathbf{r}^{\prime}\rangle=
\frac{1}{2\pi l^2}\exp\left(-\frac{(\mathbf{r} -\mathbf{r}^{\,\prime})^2}
{4l^2}-i\frac{(x+x^\prime)(y-y^\prime)}{2l^2}\right)\nonumber\\
&&\times\sum\limits_{n=0}^\infty\frac{1}{{\cal M}-(2n+1)(\hbar v^{2}_{F}/c)|eB_{\perp}|
-i(\hbar v^{2}_{F}/c)eB_{\perp}\gamma^{1}\gamma^{2}}
L_n\left(\frac{(\mathbf{r}-\mathbf{r}^{\,\prime})^2}{2l^2}\right),
\label{auxillary-prop}
\end{eqnarray}
where we integrated over $p$ by using
\begin{equation}
\int\limits_{-\infty}^\infty\,e^{-x^2}H_m(x+y)H_n(x+z)dx
=2^n\pi^{1/2}m!z^{n-m}L_m^{n-m}(-2yz),
\end{equation}
assuming $m\le n$. Here $L^{\alpha}_n$ are the generalized Laguerre polynomials, and
$L_n \equiv L^{0}_n$. The matrix $i(\hbar v_F^{2}/c)eB_{\perp} \gamma^1\gamma^2$ has eigenvalues
$\pm \hbar v_F^{2}|eB_{\perp}|/c=\pm \epsilon_B^2/2$. Therefore, we have
\begin{equation}
\frac{L_{n}(\xi)}{{\cal M}-(2n+1)(\hbar v^{2}_{F}/c)|eB_{\perp}|-i(\hbar v^{2}_{F}/c) eB_{\perp}\gamma^{1}\gamma^{2}}
=\frac{P_{-}L_{n}(\xi)}{{\cal M}-n\epsilon_B^2}+
\frac{P_{+}L_{n}(\xi)}{{\cal M}-(n+1)\epsilon_B^2},
\label{A9}
\end{equation}
where $\xi=(\mathbf{r}-\mathbf{r}^{\,\prime})^2/(2l^2)$ and the projectors $P_{\pm}$ are
\begin{equation}
P_{\pm}=\frac{1}{2}\left[1\pm i\gamma^{1}\gamma^{2}\mbox{sign}(eB_{\perp})\right].
\label{projectors}
\end{equation}
By redefining $n\to n-1$ in the second term in Eq.~(\ref{A9}),
Eq.(\ref{auxillary-prop}) can be rewritten as
\begin{eqnarray}
\langle\mathbf{r}|[{\cal M}-v_F^{2}\bm{\pi}^{2}-ieB_{\perp}(\hbar v_F^{2}/c)\gamma^1\gamma^2]^{-1}|
\mathbf{r}^{\prime}\rangle=\frac{1}{2\pi l^{2}}e^{i\Phi(\mathbf{r},\mathbf{r}^{\prime})}
e^{-\xi/2}\sum\limits_{n=0}^\infty\frac{P_{-}L_{n}(\xi)+P_{+}L_{n-1}(\xi)}{{\cal M}-
n\epsilon_B^2},
\end{eqnarray}
where $L_{-1}\equiv 0$  by definition and
\begin{equation}
\Phi(\mathbf{r},\mathbf{r}^{\prime})=-\frac{(x+x^\prime)(y-y^\prime)}{2l^2}=
-\frac{e}{\hbar c}\int\limits_{\mathbf{r}^{\prime}}^{\mathbf{r}}dz_{i}A_{i}(z)
\label{Schwinger-phase}
\end{equation}
is the Schwinger phase \cite{Schwinger} which appears due to the noncommutative character of 
magnetic translations \cite{Zak}. Since
\begin{eqnarray}
\pi_{x}e^{i\Phi} = e^{i\Phi}\hbar\left(-i\partial_{x}-\frac{y-y^{\prime}}{2l^{2}}\right),\quad
\pi_{y}e^{i\Phi} = e^{i\Phi}\hbar\left(-i\partial_{y}+\frac{x-x^{\prime}}{2l^{2}}\right),
\label{Schwinger-phase-derivities}
\end{eqnarray}
propagator (\ref{propagator1}) can be presented as a product of the phase factor and a translation
invariant part $\bar{G}(\omega;\mathbf{r}-\mathbf{r}^{\prime})$,
\begin{equation}
G(\omega;\mathbf{r},\mathbf{r}^{\prime})=e^{i\Phi(\mathbf{r},\mathbf{r}^{\prime})}
\bar{G}(\omega;\mathbf{r}-\mathbf{r}^{\prime}),
\label{full-propagator-two-parts}
\end{equation}
where
\begin{eqnarray}
\bar{G}(\omega;\mathbf{r}-\mathbf{r}^{\prime})=i\left[W-\hbar v_{F}\gamma^{1}
\left(-i\partial_{x}-\frac{y-y^{\prime}}{2l^{2}}\right)-\hbar v_{F}\gamma^{2}
\left(-i\partial_{y}+\frac{x-x^{\prime}}{2l^{2}}\right)\right]\frac{e^{-\xi/2}}{2\pi l^{2}}
\sum\limits_{n=0}^\infty\frac{P_{-}L_{n}(\xi)+P_{+}L_{n-1}(\xi)}{{\cal M}-
n\epsilon_B^2}.
\label{propagatorG-FT}
\end{eqnarray}
The Fourier transform of the translation invariant part of propagator (\ref{propagatorG-FT})
can be evaluated by performing the integration over the angle,
\begin{equation}
\int\limits_{0}^{2\pi}d\theta e^{ikr\cos\theta}=2\pi J_{0}(kr),
\end{equation}
where $J_{0}(x)$ is the Bessel function, and then using the
following formula:
\begin{equation}
 \int_0^\infty
xe^{-\frac{1}{2}\alpha x^2}L_n\left(\frac{1}{2}\beta
x^2\right)J_0(xy)dx
=\frac{(\alpha-\beta)^n}{\alpha^{n+1}}e^{-\frac{1}{2\alpha}
y^2}L_n\left( \frac{\beta y^2}{2\alpha(\beta-\alpha)}\right),
\end{equation}
valid for $y>0$ and $\mbox{Re}\,\alpha>0$. We obtain
\begin{equation}
\bar{G}(\omega,\mathbf{k}) = ie^{-k^2 l^{2}}\sum_{n=0}^{\infty}
\frac{(-1)^nD_{n}(\omega,\mathbf{k})}{{\cal M}-n\epsilon_B^2},
\label{Dn-new}
\end{equation}
with
\begin{eqnarray}
D_{n}(\omega,\mathbf{k}) = 2W\left[P_{-}L_n\left(2 k^2 l^{2}\right)
-P_{+}L_{n-1}\left(2 k^2 l^{2}\right)\right]
 + 4\hbar v_F(\mathbf{k}\cdot\bm{\gamma}) L_{n-1}^1\left(2 k^2 l^{2}\right),\,
L_{-1}^\alpha \equiv 0,
\label{Dn}
\end{eqnarray}
describing the $n$th Landau level contribution.

\section{Solutions of gap equation in purely magnetic or pseudomagnetic field}
\label{limiting-cases}

\subsection{Purely magnetic field}
\label{magnetic-LLL}

Let us find solutions of the gap equation (\ref{eq:gapEquation}) in the LLL approximation in
the case where only magnetic field is present. In this case, the gap equation takes the form:
\begin{eqnarray}
 - d_\nu(s_0) \sigma_\nu + s_0 \epsilon_Z \sigma_z =&-&\frac{1}{2}(u_0+u_z) \bigg[
\frac{d_i(s_0)\sigma_i}{E(s_0)}\theta(E(s_0)-|d_0(s_0)|)+\;\text{sgn}(d_0(s_0))\theta(|d_0(s_0)|-E(s_0))
\bigg]\nonumber\\
&+&u_\bot \bigg[\frac{d_i(-s_0)\sigma_i}{E(-s_0)}\theta(E(-s_0)-|d_0(-s_0)|)
+\;\text{sgn}(d_0(-s_0))\theta(|d_0(-s_0)|-E(-s_0))
\bigg]\nonumber\\
 & +&u_z \sum_{s'_0}\text{sgn}(d_0(s'_0)) \theta(|d_0(s'_0)|-E(s'_0)),
\label{gap-equation-LLL}
\end{eqnarray}
where $d_\nu = \tilde{\Delta}_\nu + s_0 \Delta_\nu$ and $E = \sqrt{d_x^2+d_z^2}$.
Projecting on the Pauli matrices, we get
\begin{equation}
d_0(s_0)= \frac{1}{2} (u_0-u_z) \,\text{sgn}(d_0(s_0))\,\theta(|d_0(s_0)|-E(s_0))
-(u_\bot+u_z)\,\text{sgn}(d_0(-s_0))\,\theta(|d_0(-s_0)|-E(-s_0)),
\label{d-0-equation}
\end{equation}
\begin{equation}
d_x(s_0) = \frac{1}{2}(u_0+u_z)\frac{d_x(s_0)}{E(s_0)}\theta(E(s_0)-|d_0(s_0)|)
-u_\bot \frac{d_x(-s_0)}{E(-s_0)}\theta(E(-s_0)-|d_0(-s_0)|),
\label{d-x-equation}
\end{equation}
\begin{equation}
d_z(s_0) -s_0 \epsilon_Z= \frac{1}{2}(u_0+u_z)\frac{d_z(s_0)}{E(s_0)}\theta(E(s_0)-|d_0(s_0)|)
-u_\bot \frac{d_z(-s_0)}{E(-s_0)}\theta(E(-s_0)-|d_0(-s_0)|).
\label{d-z-equation}
\end{equation}
These equations define a system of non-linear equations because $E$  is a non-linear function
of $d_x$ and $d_z$. Since $G_{int}$ approximates the Coulomb interaction, which is the strongest electron-electron interaction in graphene, we
assume in what follows that
$u_0\gg\epsilon_Z,u_z,u_\bot$. The system of equations (\ref{d-0-equation})-(\ref{d-z-equation})
has the following solutions which are consistent with the charge neutrality condition for
both $\pm$ signs:

\begin{itemize}

\item[(i)] F solution:
\begin{equation}
\tilde\Delta_0=\Delta_0=\tilde\Delta_x=\Delta_x=0, \qquad
\tilde{\Delta}_z=0, \qquad \Delta_z=\epsilon_Z\pm\frac{1}{2}(u_0+u_z+2u_\bot).
\label{ferromagnetic-LLL-1}
\end{equation}

\item[(ii)] AF solution:
\begin{equation}
\tilde\Delta_0=\Delta_0=\tilde\Delta_x=\Delta_x=0, \qquad
\tilde{\Delta}_z=\pm\frac{1}{2}(u_0+u_z-2u_\bot), \qquad
\Delta_z=\epsilon_Z,
\label{ferromagnetic-LLL-2}
\end{equation}
which exists for $|\tilde{\Delta}_z|>\epsilon_Z$ that is satisfied automatically since
we assumed $u_0\gg\epsilon_Z, u_z,u_\bot$.
\item[(iii)] CAF solution:
\begin{equation}
\begin{gathered}
\tilde\Delta_0=\Delta_0=0, \qquad
\Delta_x=\tilde{\Delta}_z=0, \qquad
\cos\theta=-\frac{\epsilon_Z}{2u_\bot}, \\
\tilde{\Delta}_x=\pm\frac{1}{2}(u_0+u_z-2u_\bot)\sin\theta, \qquad
\Delta_z=\frac{1}{2}(u_0+u_z-2u_\bot)\cos\theta.
\end{gathered}
\label{CAF-solution}
\end{equation}
The CAF solution exists only for $\epsilon_Z<2|u_\bot|$.

\item[(iv)] CDW solution:
\begin{equation}
\Delta_0=\Delta_x=\tilde{\Delta}_x=\tilde{\Delta}_z=0, \qquad
\Delta_z=\epsilon_Z, \qquad
\tilde{\Delta}_0=\pm\frac{1}{2}(u_0-3u_z-2u_\bot).
\end{equation}
\end{itemize}
The ferromagnetic and CAF solutions reproduce solutions obtained in Refs.[\onlinecite{Kharitonov,Kharitonov-edge-states}]. Note that here like in Sec.\ref{phase-diagram}
we follow Kharitonov \cite{Kharitonov,Kharitonov-edge-states} and assume that $u_\bot<0$.
In this case, the CAF solution is the ground state of the system for sufficiently small
parallel magnetic field.

Since in the present paper we consider the state with zero
filling factor, $\nu=0$, we need to control the carrier density $\rho$ which according to
Eqs.(\ref{carrier_density}),(\ref{G(s0,s12)}) in the LLL approximation  is given by the expression
\begin{equation}
\rho=\frac{1}{2\pi l^2}\sum_{s_0}{\rm sgn}\left(\mu_L(s_0)\right)\theta\left(|\mu_L(s_0)|-E(s_0)\right),
\end{equation}
where $\mu_L(s_0)=-(s_0\tilde\Delta_0+\Delta_0)$ and $E(s_0)=\sqrt{(\tilde\Delta_x+s_0\Delta_x)^2
+(\tilde\Delta_z+s_0\Delta_z)^2}$. It is easy to check that for all obtained solutions the carrier
density $\rho$ vanishes so that the filling $\nu=0$ is realized. The ground state is determined by
the solution with the lowest free energy density. The case of purely magnetic field corresponds to
the line $B_5=0$ in Figs.1 and 2 in Sec.\ref{phase-diagram}. One can see that there is a phase
transition from the CAF state to the F state as parallel magnetic field $B_{||}$ increases that
agrees with the results obtained by Kharitonov in Refs.[\onlinecite{Kharitonov,Kharitonov-edge-states}].

\subsection{Purely pseudomagnetic field}
\label{pseudomagnetic-LLL}

In this subsection, we consider solutions of the gap equation (\ref{eq:gapEquationExact-B5})
in the case where only pseudomagnetic field is present ($B_\perp=B_{||}=0$) retaining
only the LLL contribution. Without loss of generality we can
put $\mu_\nu=\tilde\mu_\nu=0$ so that $m_\nu=0$, and we are left only with parameters $d_\nu$.
Moreover, due to the absence of the Zeeman term the gap equation possesses the SU(2) spin symmetry,
so that we can take $d_x=0$. Therefore, we get the following equation for the parameters $d_0$ and 
$d_z$:
\begin{eqnarray}
 - d_z(s_{12}) \sigma_z - d_0(s_{12})
 =&-&\frac{1}{2}(u_{05}+u_{z5}) \bigg[
\text{sgn}(d_z(s_{12}))\theta(E(s_{12})-|d_0(s_{12})|)\sigma_z
+\text{sgn}(d_0(s_{12}))\theta(|d_0(s_{12})|-E(s_{12}))\bigg]\nonumber\\
&-&2u_5 \bigg[
\text{sgn}(d_z(-s_{12}))\theta(E(-s_{12})-|d_0(-s_{12})|)\sigma_z
+\text{sgn}(d_0(-s_{12}))\theta(|d_0(-s_{12})|-E(-s_{12}))
\bigg]\nonumber\\
  &+&u_{z5} \sum_{s'_{12}}\text{sgn}(d_0(s'_{12})) \theta(|d_0(s'_{12})|-E(s'_{12})),
\label{gap-equation-LLL-B5}
\end{eqnarray}
where  $d_{0,z} = \tilde{\Delta}_{0,z} - s_{12} \Delta_{0,z}$ and $E=|d_z|$.

Assuming that $u_{05}>> u_{z5},u_5$, we find the following solutions:
\begin{itemize}

\item[(i)] F solution:
\begin{equation}
\tilde\Delta_0=\Delta_0=0, \qquad
\tilde{\Delta}_z=0, \qquad \Delta_z=\pm\frac{1}{2}(u_{05}+u_{z5}-4u_5);
\label{ferromagnetic-LLL-B5-1}
\end{equation}

\item[(ii)] AF solution:
\begin{equation}
\tilde\Delta_0=\Delta_0=0, \qquad
\tilde{\Delta}_z=\pm\frac{1}{2}(u_{05}+u_{z5}+4u_5), \qquad \Delta_z=0;
\label{ferromagnetic-LLL-B5-2}
\end{equation}

\item[(iii)] QAH solution:
\begin{equation}
\tilde\Delta_z=\Delta_z=0, \qquad\tilde{\Delta}_0=0, \qquad
\Delta_0=\pm\frac{1}{2}(u_{05}+u_{z5}-4u_5).
\label{QAH-pseudomagnetic}
\end{equation}
\end{itemize}
Note that the CAF solution is no longer present due to the absence of the Zeeman term.
For the carrier density we obtain the following expression:
\begin{equation}
\rho=\frac{1}{2\pi l_5^2}\sum_{s_{12}}{\rm sgn}\left(\tilde\Delta_0-s_{12}\Delta_0\right)
\theta\left(|\tilde\Delta_0-s_{12}\Delta_0|-|\tilde\Delta_z-s_{12}\Delta_z|\right).
\end{equation}
One can check that the condition $\rho=0$ is satisfied for all obtained solutions.
The ground state of the system is determined as the solution with the lowest free energy density
(see Sec.\ref{phase-diagram}). By making use of the energy density (\ref{free-energy-density}) for $B_{\perp}=\epsilon_Z=0$, we find that the F and QAH solutions for $g_{\bot\bot}<0$ have the lowest
and equal energy densities. According to Eqs.(\ref{ferromagnetic-LLL-B5-1})
and (\ref{QAH-pseudomagnetic}), the F and QAH solutions are described by Haldane-type masses.
These results agree with the findings in Ref.[\onlinecite{Herbut-coupling}], where it was shown that pseudomagnetic field catalyses the generation of the Haldane mass. On the hand, we find that for $g_{\bot\bot}>0$ the AF state with Dirac-type mass (\ref{ferromagnetic-LLL-B5-2}) is the most preferable. Once again, like in Sec.\ref{phase-diagram}, these results show the crucial role of the coupling constant $g_{\perp\perp}$ in the selection of the ground state of the system in a strong pseudomagnetic field.

\section{Free-energy density}
\label{B}

The energy density is defined by the following expression $\Omega=-\Gamma/TV$, where $TV$ is
a space-time volume. By utilizing the Baym-Kadanoff-Jackiw-Tomboulis) formalism, we find the
effective action $\Gamma$ at its extrema in the mean-field approximation (for details see Ref.[\onlinecite{graphene-QHE}]):
\begin{equation}
\Gamma=-i \text{Tr}\left[\ln G^{-1} + \frac{1}{2}(S^{-1}G-1)\right],
\end{equation}
where the trace, logarithm, and product $S^{-1}G$ are taken in the functional sense, $G=\text{diag}(G_+,G_-)$, $1$ is the unit operator in both matrix and coordinate sense and
the expressions for the free and full propagator are given by Eqs.(\ref{inverse-free-propagator})
and (\ref{ansatz}). Performing the Fourier transform in time, integrating by parts the logarithm
term, and omitting the irrelevant surface term, we arrive at the expression (for simplicity in
this section we put constants $\hbar=c=1$):
\begin{equation}
\Gamma=-iT \int_{-\infty}^{\infty}\frac{dw}{2\pi}\text{Tr}\left[-w\frac{\partial G^{-1}(w)}{\partial w}
G(w) + \frac{1}{2}(S^{-1}(w)G(w)- 1)\right]
\end{equation}
with
\begin{equation*}
\frac{\partial G^{-1}(w)}{\partial w} = -i\gamma^0\delta(\boldsymbol{r-r'}).
\end{equation*}
The multiplier $T$ came from the functional trace in time, and now $\mbox{Tr}$ contains only spatial integration. By substituting the expression for the Green function (\ref{full-propagator-two-parts})
into $\Omega=-\Gamma/TV$, one can see that Schwinger phase $\Phi$ goes away and after the Fourier transformation we get for the energy density
\begin{eqnarray}
\Omega&= & i \int_{-\infty}^{\infty}\frac{dw}{2\pi} \int\frac{d^2k}{(2\pi)^2} \text{tr}\left[i\gamma^0w\bar{G}(w,\boldsymbol{k}) + \frac{1}{2}[-i\{(w - \epsilon_Z \sigma_Z) \gamma^0 
- v_F (\boldsymbol{k} \cdot \boldsymbol{\gamma})\}\bar{G}(w;\boldsymbol{k}) - 1]\right]\nonumber\\
&=& -\int_{-\infty}^{\infty}\frac{dw}{4\pi} \int\frac{d^2k}{(2\pi)^2} \text{tr}\{[(w+\epsilon_Z \sigma_Z)\gamma^0+v_F(\boldsymbol{k} \cdot \boldsymbol{\gamma})]\bar{G}(w;\boldsymbol{k})+i\}.
\end{eqnarray}

By making use of the explicit form of the propagator, we calculate the following integrals which 
contribute to the energy density:
\begin{eqnarray}
&&\int\frac{d^2k}{(2\pi)^2}\gamma^0 \bar{G}_s(w;\boldsymbol{k})=\frac{i|eB|}{2\pi}\sum_{n=0}^{\infty}
\gamma^0 W \frac{P_-+P_+ \theta(n-1/2)}{{\cal M}-n\epsilon_B^2},\\
&&\int\frac{d^2k}{(2\pi)^2}v_F (\boldsymbol{k} \cdot \boldsymbol{\gamma}) \bar{G}_s(w;\boldsymbol{k})=
\frac{i|eB|}{2\pi}\sum_{n=0}^{\infty}
\frac{n\epsilon_B^2}{{\cal M}-n\epsilon_B^2},
\end{eqnarray}
where in strained case we use the effective magnetic field $B=B_\bot-s_0 s_{12} B_5$ in the Landau 
energy $\epsilon_B$ and projectors $P_\pm$.

By dropping an infinite divergent term independent of the physical parameters and normalizing
$\Omega$ by subtracting its value at $m_\mu=d_\mu=\epsilon_Z=0$, we obtain the following expression:
\begin{align}
\Omega=&-\frac{i|eB|}{4\pi}
\int_{-\infty}^{\infty}\frac{dw}{2\pi} \hspace{1mm}\text{tr} \sum_{n=0}^{\infty}
\left[\frac{(w+\epsilon_Z\sigma_Z)\gamma^0 W (P_-+P_+ \theta(n-1/2)) + n\epsilon_B^2}{{\cal M}-n\epsilon_B^2}
-\frac{w^2(P_-+P_+ \theta(n-1/2)) + n\epsilon_B^2}{w^2-n\epsilon_B^2}\right].
\end{align}

After integrating this expression over frequency and taking trace, we finally arrive at
Eq.(\ref{free-energy-density}).

\end{document}